\documentclass[twocolumn,showpacs,preprintnumbers,superscriptaddress,aps,prd,nofootinbib]{revtex4-2}
\usepackage[english]{babel}
\usepackage[utf8]{inputenc}
\usepackage[colorinlistoftodos, color=green!40, prependcaption]{todonotes}
\usepackage{amsthm}
\usepackage{lipsum}
\usepackage{cancel}
\usepackage{mathtools}
\usepackage{physics}
\usepackage{xcolor}
\usepackage{graphicx}
\usepackage[left=23mm,right=13mm,top=35mm,columnsep=15pt]{geometry} 
\usepackage{adjustbox}
\usepackage{placeins}
\usepackage[T1]{fontenc}
\usepackage{lipsum}
\usepackage{csquotes}
\usepackage[pdftex, pdftitle={Article}, pdfauthor={Author}]{hyperref} 
\usepackage{booktabs}
\usepackage{multirow}
\usepackage{ulem}
\usepackage{hyperref}
\hypersetup{
  colorlinks  = true,
  urlcolor    = blue,
  linkcolor   = red,
  citecolor   = blue
}

\begin{document}
\title{Super-exponential Primordial Black Hole Production\\ via Delayed Vacuum Decay}

\author{Yanda Wu}
    \email{yanda.wu7@sjtu.edu.cn}
    \affiliation{Tsung-Dao Lee Institute,
Shanghai Jiao Tong University, Shanghai 201210, China}
\affiliation{Shanghai Key Laboratory for Particle Physics and Cosmology,
Key Laboratory for Particle Astrophysics \& Cosmology (MOE),
Shanghai Jiao Tong University,
Shanghai 200240, China}

\author{Stefano Profumo}
    \email{profumo@ucsc.edu}
    \affiliation{Department of Physics, University of California, Santa Cruz (UCSC),
Santa Cruz, CA 95064, USA}
\affiliation{Santa Cruz Institute for Particle Physics (SCIPP),
Santa Cruz, CA 95064, USA}

\begin{abstract}
If a cosmological first-order phase transition occurs sufficiently slowly, delayed vacuum decay may lead to the formation of primordial black holes. Here we consider a simple model as a case study of how the abundance of the produced black holes depends on the model's input parameters. We demonstrate, using both numerical and analytical arguments and methods, that the black hole abundance is controlled by  a double, ``super''-exponential dependence on the three-dimensional Euclidean action over temperature at its minimal value. We show that a modified expansion rate during the phase transition, such as one driven by an additional energy density component, leads to a weaker dependence on the underlying model parameters, but maintains the same super-exponential structure. We argue that our findings generalize to any framework of black hole production via delayed vacuum decay.
\end{abstract}

\maketitle

\section{Introduction} \label{sec:outline} 

Primordial black holes (PBHs) are potential messengers not only of the early universe's structure, but also compelling candidates for the cosmological dark matter \cite{escriva, carr}. Unlike stellar black holes, PBHs could spawn from density fluctuations in the early universe, offering a unique probe into physics beyond the Standard Model and conditions of the very early universe. In particular, first-order phase transitions (FOPTs) have been identified as a natural framework for PBH formation, with recent attention directed towards scenarios involving delayed vacuum decay during such transitions \cite{Liu:2021svg,ref1,ref2}. 

In the early universe, strongly first-order phase transitions induce nucleation of bubbles of the ``true'', energetically-favored vacuum, which then expand and collide in the background of the false, unbroken phase. If these collisions result in regions of sufficiently high energy densities, gravitational collapse may ensue, culminating in black hole formation \cite{ref5, Khlopov:1998nm}. This paper focuses, instead,  on the alternate possibility that delayed vacuum decay, wherein a metastable vacuum state persists longer than typically anticipated, can enhance its energy density relative to the surrounding regions, leading to gravitational collapse and  PBH production \cite{Liu:2021svg,ref1}. Ref.~\cite{Liu:2021svg} first proposed the general mechanism discussed here, as a natural consequence of quantum tunneling randomness. The paper established the foundational idea that postponed vacuum decay in certain Hubble volumes creates overdense regions that eventually collapse into PBHs, predicting a nearly monochromatic mass function and connecting PBH production to observable gravitational wave signatures.

Recent theoretical models have expanded on this general mechanism. Kawana et al. \cite{ref1} elaborate on a scenario where inhomogeneities in vacuum energy decay during a first-order phase transition seed over-densities that precipitate PBH formation. They subsequently raised significant concerns about the quantitative derivation of the PBH abundance fraction $f_{\rm pbh}$ in earlier studies. Their critique focused on the treatment of statistical variations in bubble nucleation histories and how this affects the calculation of overdensities that eventually form PBHs. They developed a more rigorous analytical approximation for the probability of a patch to collapse into a PBH as a function of the phase transition duration, showing that first-order phase transitions taking more than 15\% of a Hubble time to complete can produce observable PBH populations. Here, we augment Ref.~\cite{ref1}'s results  by offering a more complete mathematical description, that also accommodates modified expansion rates during phase transitions, demonstrating how the fundamental super-exponential structure is maintained even when parameter dependencies are weakened. In a related study, Flores et al. \cite{ref2} revisit supercooled FOPTs and argue that gradient energy in the bubble walls could in itself lead to PBH formation, effectively showcasing how deviations from equilibrium states can catalyze phenomena in the early universe such as PBH formation.

Theoretical predictions are further sharpened by detailed numerical simulations that analyze critical parameters such as energy scales and phase transition dynamics. Jedamzik and Niemeyer \cite{ref5} conducted simulations indicating that latent heat associated with first-order transitions reduces the threshold for PBH formation from density fluctuations. Moreover, Goncalves et al. \cite{Goncalves:2024vkj} examined PBH formation in the context of a gauge singlet extension to the Standard Model, linking the PBH mass distribution to the electroweak scale and emphasizing the observable signatures for experiments like Laser Interferometer Space Antenna and microlensing surveys.

The implications of the processes discussed above are far-reaching: PBHs could potentially contribute to the cosmological dark matter, and influence cosmic microwave background anisotropies \cite{escriva,carr}.  PBHs could also serve as seeds for vacuum decay in theories involving metastable states, as demonstrated in studies connecting this decay to PBH-induced true vacuum bubbles \cite{ref4}. The potential detectability of gravitational waves from these phase transitions presents an exciting and complementary opportunity to observe early universe events directly \cite{Wang:2020jrd}.

Despite significant advances in this area, challenges remain, particularly in accurately predicting PBH mass spectra and the abundance of the produced PBHs. Numerical simulations and phenomenological models will need to incorporate complex interactions during phase transitions \cite{Khlopov:1999ys}. Furthermore, reconciling theoretical predictions with numerical results and with stringent observational constraints, such as those from microlensing surveys, is essential for refining the relevant parameter spaces \cite{Goncalves:2024vkj, Baker:2021sno}.

The present study aims to explore, analytically, the conditions under which delayed vacuum decay leads to PBH formation during a strongly first-order phase transition in a simplified model consisting of a single scalar field driving the phase transition. In particular, our principal goal is to derive an {\it analytical understanding} of how the microphysical parameters entering the potential driving the phase transition impact the resulting PBH abundance.

\section{PBH from delayed vacuum decay} \label{sec:develop}

A first order phase transition proceeds via bubble nucleation of the ``true'' vacuum of the broken phase inside the ``false'' vacuum of the unbroken phase. The nucleation time and nucleation temperature, however, generally differ in different spacetime patches. Suppose a patch $P$ nucleates after its surrounding background patches (denoted by $B$): the energy density (radiation plus vacuum energies) of  patch $P$ would then be larger than that of $B$. If the energy density ratio in patches $P$ and $B$ exceeds a certain density contrast limit, driven by the larger contribution of vacuum energy density in the delay-decay patches $P$, a generic expectation is that the $P$ regions potentially collapse into  black holes. 

Here, we use a simplified model involving a single real scalar field, uncharged under the gauge symmetry groups of the Standard Model, to study this mechanism analytically, and validating our findings numerically. The effective, finite-temperature potential of the model we consider follows the one considered in Ref.~\cite{Kanemura:2024pae},
\begin{equation}\label{eq:potential}
 \begin{aligned}
V(\phi,T)&=\frac{1}{2}\left( \frac{\mu_3 \omega - m_\phi^2}{2}+c T^2\right)\phi^2 -\frac{\mu_3}{3}\phi^3 \\
&\quad + \frac{m_\phi^2}{8\omega^2} \left( 1+\frac{\mu_3 \omega}{m_\phi^2}\right)\phi^4,
\end{aligned}   
\end{equation}
The potential is thus described by four independent parameters: $m_\phi$, $\omega$, $c$, and $\mu_3$. We discuss additional details of this model in the Appendix. 

The thermal tunneling probability between the true and false vacua is given, as usual \cite{Callan:1977pt,Coleman:1980aw}, by 
\begin{align} \label{eq:nucleation_rate}
    \Gamma \simeq T^4 \left(\frac{S_3(T)}{2\pi T}\right)^{3/2} e^{-S_3(T)/T},
\end{align}
where $S_3$ is the three dimensional Euclidean action. 

The computation of the PBH relic abundance proceeds as follows \cite{Kanemura:2024pae}:
We consider two types of Hubble patches, ``normal'' ones and ``delayed-decay'' ones. The normal and delayed-decay patches are characterized by different radiation and vacuum energy densities, which evolve radically differently with time after the phase transition's critical time $t=t_c$. The vacuum energy density can be cast as
\begin{align}
    \rho_v = F(t) \Delta V(t),
\end{align}
where $F(t)$ is the false vacuum volume fraction and $\Delta V(t)$  the difference in the effective potential values between the true and false vacua. 

Note that for the normal patch $F(t)=1$ when $t\leq t_c$ and $F(t)<1$ when $t>t_c$, while for the delayed-decay patch $F(t)=1$ when $t\leq t_c+\Delta t \equiv t_d$  and $F(t)<1$ when $t>t_d$, where naturally $\Delta t$ indicates the characteristic time delay of nucleation in the two patches. For the delayed-decay patch at $t>t_d$, $F(t)$ is given by
\begin{align} \label{eq:Ft_delayed_patch}
F(t)=\exp \left\{-\frac{4\pi}{3}\int_{t_d}^t d t^\prime \Gamma(t^\prime) a^3(t^\prime) \gamma^3(t^\prime,t) \right\},
\end{align}
where
\begin{align}
    \gamma(t^\prime,t)= \int_{t^\prime}^t d t^{\prime \prime}\frac{v_\omega}{a(t^{\prime \prime})},
\end{align}
with $v_\omega$  the bubble wall velocity. The equation above also gives the evolution of $F(t)$ for the normal patch when $t>t_c$, with the replacement of $t_d \rightarrow t_c$ in the time integral.

The dynamical evolution of the radiation energy in both patches cannot be written in a closed form, unlike the vacuum energy, and it therefore requires a numerical solution to the first and second Friedmann equations. 

We label the energy density in the normal patch with $\rho_\text{nor}$ and that in the delayed-decay patch as $\rho_\text{del}$. When the density ratio between $\rho_\text{del}$ and $\rho_\text{nor}$ exceeds a certain critical value $\delta_c$, i.e.\footnote{Flores et al. \cite{ref2} have cast doubt on this energy critical collapse criteria for forming PBHs. However, a recent visit by Cai et al. \cite{Cai:2024nln} clarifies that the original critical collapse criterion still applies, as in  inflationary PBH production scenarios. We postpone a full discussion of this issue to future work.}
\begin{align}
\delta(t)=\frac{\rho_\text{del}(t)}{\rho_\text{nor}(t)}-1>\delta_c,
\end{align}
gravitational collapse occurs, and we posit, as natural, that a PBH is formed.

After $t_d$, i.e. after the ``critical delay time" for the delayed-decay patch, suppose $\delta(t)$ increases and reaches its maximum value $\delta_{\text{max}}$ at $t=t_{\text{pbh}}$. Such value $\delta_{\text{max}}$ may be (i) smaller, (ii) equal, or (iii) larger than $\delta_c$. Only cases (ii) and (iii) allow the delayed-decay patch to collapse into a PBH. As suggested by Ref.\cite{Kanemura:2024pae}, the dominant PBH production condition is the threshold condition $\delta_{\text{max}}(t)=\delta_c$, from which we can determine $t_d$ and $t_{\text{pbh}}$ simultaneously. Larger value of $\delta_{\text{max}}(t)$ requires larger value of $t_d$, which causes the $F(t)$ and number of delayed-decay patches to decrease rapidly, resulting in the abundance of PBH to also rapidly decrease. Here, for definiteness, we take $\delta_c=0.45$ as the PBH formation criterion (see e.g.  \cite{carr, escriva}). 

The masses of the PBH resulting from delayed vacuum decay read
\begin{align}
m_{\text{pbh}}\simeq \gamma V_{H,\text{del}}(t_{\text{pbh}})\rho_{\text{del}}(t_{\text{pbh}}),
\end{align}
where $V_{H,\text{del}}(t_{\text{pbh}})=(4\pi/3)H_{\text{del}}^{-3}(t_{\text{pbh}})$ is the delayed-patch Hubble volume. $\gamma$ is a number of ${\cal O}(1)$ referring to the ratio between the PBH mass and the Hubble mass \cite{carr}. The PBH energy density $\rho_{\text{pbh}}$ reads
\begin{align} \label{eq:rho_pbh_definition}
    \rho_{\text{pbh}} \simeq P(t_d) \frac{m_{\text{pbh}}}{V_{H,\text{nor}}},
\end{align}
where
\begin{equation}\label{eq:Pint_definition}
 \begin{aligned}
	P(t_d)&=\exp\left\{-\frac{4\pi}{3}\int_{t_c}^{t_d}dt\ \mathcal{N}(t_{\text{pbh}}) a_{\text{del}}^3(t) \Gamma_{\text{del}}(t) \right\},
\end{aligned}   
\end{equation}
is the probability of no bubble nucleation between $t_c$ and $t_d$ and $\mathcal{N}(t_{\text{pbh}})=a_{\text{del}}^{-3}(t_{\text{pbh}})H_{\text{del}}^{-3}(t_{\text{pbh}})$. 

The final PBH relic abundance $\Omega_{\text{pbh}}=\rho^0_{\text{pbh}}/\rho_0$, where $\rho^0_{\text{pbh}}=\rho_{\text{pbh}} s_0/s$ and, as usual, $s\simeq 2\pi^2 g_{*} T^3_{\text{nor}}(t_{\text{pbh}})/45$.\footnote{The critical energy density and entropy are defined as: $\rho_0=(3M_{\text{pl}}^2/8\pi)H_0^2$ with $M_{\text{pl}}$  the Plank Mass and $H_0$  the current Hubble rate; $s_0\simeq 2891.2 \text{cm}^{-3}$ is the entropy today \cite{ParticleDataGroup:2024cfk}.} 
It is customary to also define the  fraction of dark matter consisting of PBH, $f_{\text{pbh}}$, as
\begin{align}	\label{eq:f_pbh_definition}f_{\text{pbh}}=\frac{\rho^0_{\rm pbh}}{\rho_{\rm dm}}=\frac{\Omega_{\text{pbh}}}{\Omega_{\text{dm}}},
\end{align}
where $\rho_{\text{dm}}h^2$ is the dark matter cosmological abundance, and $\Omega_X=\rho_X/\rho_{\rm crit}$. 

\subsection{Numerical calculation of $f_{\text{pbh}}$}

\begin{table}[t]
    \centering
    \caption{Details of the five BMs. In all BMs, we take $m_\phi=300$ MeV, and the $\mu_3^*$ is chosen for $f_\text{pbh}=1$. The subscript 1 refers to the parameter value for BM1 and, for BM2 to BM5, we list the values of model parameters in terms of relative changes with respect to BM1. The final column shows the ``strength'' of the phase transition $v_c/T_c$, where $v_c$ being the vacuum expectation value at the critical temperature $T_c$.
    }
    \begin{tabular}{|lcccc|}
        \toprule
         & $\boldsymbol{\omega}$ (MeV) & $\boldsymbol{c}$ & $\boldsymbol{\mu_3^*}$ (MeV) & $\boldsymbol{v_c/T_c}$ \\ \midrule
        \textbf{BM1:} & $\omega_1=$860 & $c_1=$0.140 & 160.203 & 2.26 \\ \hline
        \textbf{BM2:} & $\omega_1+34.29$ & $c_1-0.017$ & $153.659$ & 2.20 \\ \hline
        \textbf{BM3:} & $\omega_1+34.29$ & $c_1-0.06$ & $160.066$ & 1.85 \\ \hline
        \textbf{BM4:} & $\omega_1+11.43$ & $c_1-0.026$ & $160.419$  & 2.10 \\ \hline
        \textbf{BM5:} & $\omega_1+80$ & $c_1-0.009$ & $142.557$ & 2.33 \\ \bottomrule
    \end{tabular}
\label{tab:BM_values_in_fpbh_u3_u3star}
\end{table}

\begin{figure}[t]
  \centering
\includegraphics[width=0.5\textwidth]{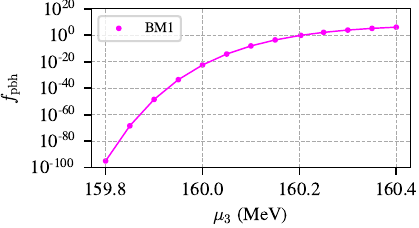}\\%
\includegraphics[width=0.5\textwidth]{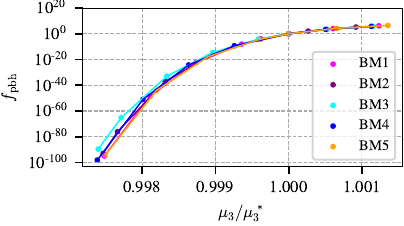}%
  \caption{ $f_{\text{pbh}}$ with $\mu_3$ under BMs given in Table.\ref{tab:BM_values_in_fpbh_u3_u3star}. Upper figure: $f_{\text{pbh}}$ with $\mu_3$ in BM1; Bottom figure:
  $f_{\text{pbh}}$ with $\mu_3/\mu_3^*$ in five BMs, where $\mu_3^*$ is chosen when $f_{\text{pbh}}=1$ within each BM.   
}
\label{fig:fpbh_mu3_mu3star}
\end{figure}

In the simplified model introduced above, and following the aforementioned delayed vacuum decay mechanism for PBH production, we solve numerically for the PBH-to-dark matter relative abundance $f_{\text{pbh}}$. \footnote{Note that our numerical results refer here to the numerical solutions of the differential equations or integrals involved in $f_\text{pbh}$, rather than actual numerical
simulations of PBH collapse in general relativity or lattice simulations.} 

For our numerical results, we first define five different {\it benchmark models} (BM), featuring different parameter inputs for the potential in Eq.~\eqref{eq:potential}  as listed in Tab.~\ref{tab:BM_values_in_fpbh_u3_u3star}. Notice that because of the extreme sensitivity of the abundance of PBH on the parameters in the effective potential, for convenience we define BM2-5 off of the values of the reference BM1. 

We show our numerical results for the abundance $f_{\rm pbh}$ as a function of the key parameter $\mu_3$ in the upper panel of  Fig.~\ref{fig:fpbh_mu3_mu3star}. The bottom panel normalizes the input value of $\mu_3$ to the value $\mu_3^*$ at which, for the given choices of the {\it other} parameters in the potential, $f_{\rm pbh}=1$. 
The lower panel provides strong evidence that the dependence on $f_\text{pbh}$ on underlying theory parameters is controlled by a universal quantity, to be explored below. 

The two panels strikingly illustrate the very strong sensitivity of the PBH abundance $f_{\rm pbh}$ on the underlying potential input parameters, here the cubic term coefficient $\mu_3$. The lower panel, where as explained we normalize each BM's $\mu_3$ value to the particular value $\mu_3^*$ giving $f_{\rm pbh}=1$, shows how the dependence on the cubic coupling is in fact a {\it universal} feature of the model under consideration.  
We explore below the mathematical and physical structure that underpins what is shown in the figure.

\subsection{Analytical estimate of $f_{\text{pbh}}$}
We can factorize quantities upon which $f_{\text{pbh}}$ depends into two categories: those {\it outside} and those {\it inside} the $P(t_d)$ integral, where $P(t_d)$ is the probability of no bubble nucleation defined in Eq.~(\ref{eq:Pint_definition}). 

Our strategy is as follows: we argue that quantities outside the $P(t_d)$ integral can be regarded as {\it parameter-independent constants} (we will prove this claim later); for quantities inside the $P(t_d)$, we define
\begin{align} \label{eq:Pint_true_definition}
    P_\text{int}\equiv \frac{4\pi}{3}\int_{t_c}^{t_d}dt\ \mathcal{N}(t_{\text{pbh}}) a_{\text{del}}^3(t) \Gamma_{\text{del}}(t),
\end{align}
so that $P(t_d)=\exp(-P_\text{int})$. 

\begin{figure}[t]
  \centering
\includegraphics[width=0.5\textwidth]{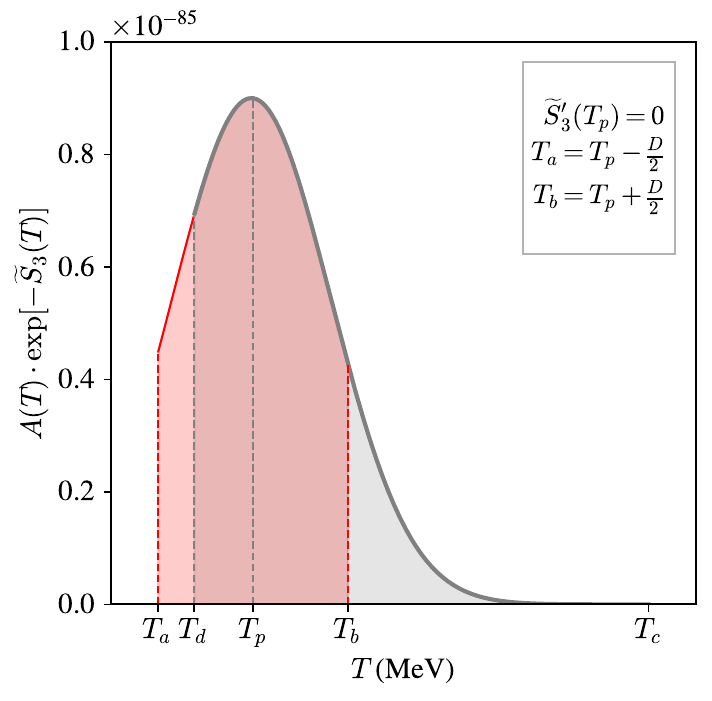}%
\caption{Approximation of the integral of $\widetilde{P}_\text{int}$, given in Eq.(\ref{eq:P_int_tilde_definition}). $T_p$ denotes the temperature when $\widetilde{S}_3(T)$ reaches its minimum. The grey shadow region represents the full integral from $T_d$ to $T_c$, while the red shadow region denotes the approximated integral from $[T_a,T_b]$, which is centered around $T_p$ with a typical full width at half maximum of the integrand, denoted by $D$.
}
\label{fig:saddle_point_appro_illus}
\end{figure}

 For the time integral from $t_c$ to $t_d$, we approximate the integral around the peak value of the $\exp(-S_3(T)/T)$ in the saddle-point approximation. We first convert the integral from time to temperature, with the standard relation, $dt=-dT/(H_\text{del}(T)T)$; we further put into the definition of the nucleation rate, $\Gamma_\text{del}$, as given in Eq.(\ref{eq:nucleation_rate}). Then, we define two quantities, $\widetilde{S}_3(T)$ and $A(T)$
\begin{align}
\label{eq:S_3oT_tilde_definition}
   \widetilde{S}_3(T)&\equiv \frac{S_3(T)}{T}, \\
  \label{eq:A_T_definition}
   A(T)&\equiv \frac{T^3}{H_\text{del}(T)}a^3_\text{del}(T)\left( \frac{\widetilde{S}_3(T)}{2\pi}\right)^{3/2}.
\end{align}
The $P_\text{int}$ can be written as
\begin{equation} \label{eq:P_int_tilde_definition}
 \begin{aligned}
P_\text{int}&=\frac{4\pi}{3}\mathcal{N}(t_{\text{pbh}})\int_{T_d}^{T_c}dT\ A(T) \exp[-\widetilde{S}_3(T)]\\
    &\equiv\frac{4\pi}{3}\mathcal{N}(t_{\text{pbh}}) \widetilde{P}_\text{int}\ \ ,
\end{aligned}   
\end{equation}
where $T_c$ and $T_d$ denote the critical temperature of the normal patch and delayed-decay patch, respectively. For the model parameters leading into successful PBH, the evolution of $A(T)\exp[-\widetilde{S}_3(T)]$ with $T$ is shown in Fig.\ref{fig:saddle_point_appro_illus}. We list the values of various quantities appearing in the $f_\text{pbh}$ approximation in Table.\ref{tab:BM_values_in_f_pbh}, according to this table and Fig.\ref{fig:saddle_point_appro_illus}, the general features for different model parameters including: (i) the difference between the peak temperature of $\exp[-\widetilde{S}_3(T)]$(denoted by $T_p$) and the peak temperature of $A(T)\exp[-\widetilde{S}_3(T)]$ is negligible; (ii) the magnitude of $T_d,T_p$ and $T_c$ obeys: $T_d<T_p \ll T_c$; (iii) the value of $\widetilde{S}_3(T_p)$ falls in the range $[171,176]$. In  Fig.\ref{fig:saddle_point_appro_illus}, we approximate the  original integral range (from $T_d$ to $T_c$) into a range $[T_a,T_b]$, which is centered around $T_p$ with full width at half maximum of the integrand, denoted by $D$. The magnitude of $D$ is of order $\sqrt{\widetilde{S}_3(T_p)}\sim 10$.\footnote{In principle, $D$ has the dimension of temperature (MeV), while $\widetilde{S}_3(T_p)$ is dimensionless.}

We then expand the $T$ around $T_p$ with the dimensionless parameter $y$,
\begin{align}
   T=T_p+D\cdot y. 
\end{align}
With the expansion of $A(T)$ and $\widetilde{S}_3(T)$ to quadratic order of $y$, the $\widetilde{P}_\text{int}$ integral then reads
\begin{equation}
\begin{aligned}
\widetilde{P}_\text{int}&\simeq D\int_{-\frac{1}{2}}^{\frac{1}{2}}dy\ \left[A(T_p)+A'(T_p)Dy+ \frac{1}{2}A''(T_p)D^2y^2 \right]\\
&\quad \quad \quad \times \exp\left[-\widetilde{S}_3(T_p)-\frac{1}{2}\widetilde{S}''_3(T_p) D^2y^2\right], \\
&\simeq A(T_p)\sqrt{\frac{2\pi}{\widetilde{S}''_3(T_p)}}\exp \left[-\widetilde{S}_3(T_p) \right](1+\delta(T_p)),
\end{aligned}
\end{equation}
where the prime on $A(T)$ or $\widetilde{S}_3(T)$ denotes the derivative with respect to $T$; where we push the integral limit of $y$ into $[-\infty,+\infty]$ in the second line. The final result does not depend explicitly on $D$. The $\delta(T_p)$ term reads:
\begin{align}
\delta(T_p)=\frac{1}{2}\frac{A''(T_p)}{A(T_p)}\frac{1}{\widetilde{S}''_3(T_p)}.
\end{align}
From Table.\ref{tab:BM_values_in_f_pbh}, we see that $\delta(T_p)\lesssim \mathcal{O}(0.01)\ll 1$, so we can approximate $\widetilde{P}_\text{int}$ as
\begin{align}
\widetilde{P}_\text{int}=A(T_p)\sqrt{\frac{2\pi}{\widetilde{S}''_3(T_p)}}\exp \left[-\widetilde{S}_3(T_p)\right].
\end{align}
We further define $\mathcal{Q}$ as
\begin{align} \label{eq:Q_definition_new}
\mathcal{Q}\equiv \frac{4\pi}{3}\mathcal{N}(t_\text{pbh})A(T_p)\sqrt{\frac{2\pi}{\widetilde{S}''_3(T_p)}},
\end{align}
as been shown in Table.\ref{tab:BM_values_in_f_pbh}, the $\mathcal{N}(t_{\text{pbh}})$ is a very large number of order $\mathcal{O}(10^{85})$, largely insensitive to  variations in the model parameters. Finally, the $P_\text{int}$ is approximated into
\begin{align}
    P_\text{int}\simeq \mathcal{Q}\exp \left[-\widetilde{S}_3(T_p)\right],
\end{align}
where the value of $\mathcal{Q}\sim 1\times 10^{77}$, whose variation has a minor impact on $f_\text{pbh}$ compared those from  $\widetilde{S}_3(T_p)$.

\begin{table}[t]
    \centering
    \caption{Relevant quantities entering the approximation of $f_\text{pbh}$ in three BMs, where we choose $m_\phi=300$ MeV. $\boldsymbol{V_{\text{H,nor}}}$ and $\mathcal{N}(t_\text{pbh})$ are in unit of (MeV$^{-3}$), and all temperatures are in unit of MeV.}
    \begin{tabular}{ccccc} 
        \toprule
         & $\boldsymbol{\omega}$ & $\boldsymbol{c}$ & $\boldsymbol{\mu_3}$ & Numerical $\boldsymbol{f_{\text{pbh}}}$ \\
         & $T_d$ & $T_p$ & $T_c$ & $\widetilde{S}_3(T_p)$ \\
         & $\widetilde{S}''_3(T_p)$ & $\delta(T_p)$ & $\mathcal{N}(t_\text{pbh})$ & $Q$ \\
         & $\boldsymbol{m_{\text{pbh}}}(g)$ & $ \boldsymbol{V_{\text{H,nor}}}$  & $ \boldsymbol{s_0/s}$ & $\boldsymbol{\mathcal{M}}$  \\ 
         & & & & Appro. $f_\text{pbh}$ \\
         \midrule
        \multirow{5}{*}{\textbf{BMa:}} & 860 & 0.14 & 160.2 & 0.64 \\ 
        & 144.2 & 147.2 & 306.5 & 174.1 \\
         & 0.020 & 0.006 & $2.1\times 10^{85}$ & $1.2\times 10^{77}$ \\
              & $2.4\times 10^{33}$   & $5.1 \times 10^{51} $   & $1.8 \times 10^{-36}$     & $4.8\times 10^7$ \\
              & & & & $2.2\times 10^{-5}$ \\
              \hline
        \multirow{5}{*}{\textbf{BMb:}} & 894.29 & 0.123 & 153.4 & 1.14$\times 10^{-37}$ \\
        & 149.5 & 157.2 & 327.9 & 172.5 \\
         & 0.017 & 0.007 & $2.0\times 10^{85}$ & $1.1\times 10^{77}$ \\
         & $1.9\times 10^{33}$   & $2.6\times 10^{51} $   & $9.8\times 10^{-37}$     & $4.1 \times 10^7$ \\ 
         & & & & $1.0\times 10^{-46}$ \\
         \hline
        \multirow{5}{*}{\textbf{BMc:}} & 940 & 0.131 & 142.2 & 4.90$\times 10^{-95}$ \\  
        & 144.8 & 153.6 & 323 & 171.6 \\
         & 0.017 & 0.006 & $1.8\times 10^{85}$ & $9.4\times 10^{76}$ \\
        & $1.7\times 10^{33}$   & $1.9\times 10^{51}$    & $8.3 \times 10^{-37}$     & $4.3\times 10^7$  \\
        & & & & $2.1\times 10^{-110}$ \\
        \bottomrule
    \end{tabular}
\label{tab:BM_values_in_f_pbh}
\end{table}

Now we turn to the quantities outside the $P_\text{int}$, which we denote as $\mathcal{M}$. According to Eq.(\ref{eq:rho_pbh_definition}) and Eq.(\ref{eq:f_pbh_definition}), $\mathcal{M}$ read
\begin{align}
 \mathcal{M} =  \frac{m_{\text{pbh}}}{V_{\text{H,nor}}} \frac{s_0}{s} \frac{1}{\rho_0} \frac{1}{\Omega_{\text{dm}}},
\end{align}
which approximately falls in the range $[4\times 10^7,5\times 10^7]$, as shown in Table.\ref{tab:BM_values_in_f_pbh}. Since $f_\text{pbh}$ is proportional to $\mathcal{M}$, the variation of its values has negligible effect to $f_\text{pbh}$, as compared to the variation from quantities inside the $P_\text{int}$. In our toy model, we will take $\mathcal{M}=4\times 10^7$ in $f_\text{pbh}$ approximations.

\begin{figure}[t]
  \centering
\includegraphics[width=0.5\textwidth]{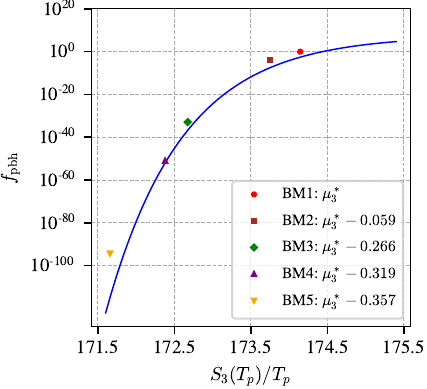}%
  \caption{$f_{\text{pbh}}$ with $S_3(T_p)/T_p$, where $T_p$ is the temperature when $S_3(T_p)/T_p$ reaches its minimum value. The solid blue line refers to our analytic approximation of $f_\text{pbh}$, given in Eq.(\ref{eq:f_pbh}), with $\mathcal{M}=4\times 10^7$ and $\mathcal{Q}=1\times 10^{77}$. The five BMs corresponding to the our numerical values of $f_{\text{pbh}}$ of model parameters given in Table.\ref{tab:BM_values_in_fpbh_u3_u3star}, where $\mu_3^*$ denotes the value of $\mu_3$ with $f_\text{pbh}=1$. For each BM, we first shift $\mu_3$ away from $\mu_3^*$, thereby generating a range of $S_3(T_p)/T_p$. We then compute the corresponding numerical values of $f_\text{pbh}$. 
}
\label{fig:fpbh_S3oT}
\end{figure}

In summary, the analysis above  shows our main finding that {\it the PBH abundance $f_{\text{pbh}}$ is  super-exponential in the Euclidean action-to-temperature ratio}, since 
\begin{align} \label{eq:f_pbh}
f_{\text{pbh}}\simeq \mathcal{M} \exp\left(-\mathcal{Q}\exp\left(-S_3(T_p)/T_p\right)\right),
\end{align}
where we put back the definition of $\widetilde{S}_3(T_p)$ in Eq.(\ref{eq:S_3oT_tilde_definition}). We now comment on the generality of our main result. A crucial step in our derivation is the saddle-point approximation of $P_\text{int}$, which yields the form, $\mathcal{Q}\exp\left(-S_3(T_p)/T_p\right)$. For a general phase transition model with delayed vacuum decay, we expect that the integrand of $P_\text{int}$ (i.e. $ A(T) \exp[-\widetilde{S}_3(T)]$) shares the same properties we described below Eq.~(\ref{eq:P_int_tilde_definition}). Hence the saddle-point approximation remains valid, and the resulting {\it super-exponential} character of $f_\text{pbh}$ persists. In our toy model, $\mathcal{Q}$ contains a factor $\mathcal{N}(t_\text{pbh})\sim \mathcal{O}(10^{85})$ and in later Sec.~\ref{sec:super_fast_universe} we will show that reducing $\mathcal{Q}$ weakens the {\it super-exponential} sensitivity of $f_\text{pbh}$ to $S_3(T_p)/T_p$\footnote{Which can be seen from Fig.~\ref{fig:f_pbh_with_phi_components_research}, where $S_3(T_p)/T_p$ increases monotonically with $\mu_3$. }. Thus a large numerical factor in $\mathcal{Q}$ is essential to our finding that $f_\text{pbh}$ is overwhelmingly more sensitive to variations in $S_3(T_p)/T_p$ than to comparable variations in $\mathcal{M}$ or $\mathcal{Q}$. If other models also feature a large, nearly parameter-independent $\mathcal{N}(t_\text{pbh})$, then variation in remaining components of $\mathcal{Q}$ or in $\mathcal{M}$ will likewise be subdominant to variations in $S_3(T_p)/T_p$. On this basis, we conclude that our expression of $f_\text{pbh}$ in Eq.~(\ref{eq:f_pbh}) and its {\it super-exponential} dependence to $S_3(T_p)/T_p$ remains valid for a broad class of phase transition models with delayed vacuum decay. 

We validate our results  plotting our estimated $f_{\text{pbh}}$ in Fig.\ref{fig:fpbh_S3oT} for fixed $\mathcal{M}$ and $\mathcal{Q}$, with the 5 benchmark models under consideration. The figure illustrates how the PBH abundance is very sensitive to small variations in the three-dimensional euclidean action to temperature ratio $S_3(T_p)/T_p$, and how our analytical results are reflected with good accuracy by our numerical computations.  

Several insights and lessons can be drawn from our key finding above: 
 Fig.\ref{fig:fpbh_S3oT} illustrates that any successful model parameters leading to considerable numbers of  PBH collapses have the feature that $S_3(T_p)/T_p$ falls in the relatively narrow region $[171.5,175.5]$. A similar statement, albeit not necessarily with the same {\it numerical} values, likely generalizes to other regions of parameter space in our toy model, as well as to entirely different models, enabling to select FOPT models that successfully form PBHs with considerable abundance.

\subsection{Super-exponential behavior of $f_\text{pbh}$ with a modified, superfast early universe expansion}\label{sec:super_fast_universe}
Our results depend critically on the background cosmology, and on the (standard) assumption that the broken phase patches' energy density redshift as radiation, i.e. that the universe evolves in a radiation-dominated background.

However, this is not necessarily the case, as observations of the Hubble rate being driven by a dominant energy density redshifting as radiation is limited, back in time/temperature, to the epoch of Big Bag Nucleosynthesis (times in the few seconds, temperatures around the MeV). It is a distinct possibility that the universe's energy density be dominated, at earlier times, by a species that redshifts ``faster'' (i.e. with a higher power of temperature, or inverse scale factor) than radiation. An example is if the universe is dominated by a fast-rolling scalar field (``kination''), with equation of state $w=P/\rho=+1$.

Here, we study how our results described above are affected by a non-standard, ``superfast'' expansion rate of the universe at early times. Specifically, we will address whether the super-exponential form of Eq.~(\ref{eq:f_pbh}) continues to hold, and what the dependence of $f_{\rm pbh}$ on the potential model parameters is.

For definiteness, we consider a new cosmological component $\phi$ in the early universe, whose energy density red-shifts as \cite{DEramo:2017gpl}
\begin{align}
    \label{eq:phi_red_shift}
    \rho_\phi\sim a^{-(4+n)}.
\end{align}
If $n>0$, the energy density of  $\phi$ is asymptotically larger than that of  radiation at early times.
We further define $\rho_R=\rho_r+\rho_\phi$, where $\rho_r$ is the usual, standard radiation energy density. In the absence of vacuum energy, we can cast $\rho_R$ at any temperature as \footnote{We assume for simplicity that the number of relativistic degrees of freedom $g_*$ remains unchanged; thus $\rho_r = \pi^2g_* T_c^4/30$, where we choose $g_*=247/4$ for temperatures around 300 MeV \cite{ParticleDataGroup:2024cfk}. }
\begin{align}
    \label{eq:rho_R_with_T}
    \rho_R=\rho_r\left[1+\left( \frac{T}{T_r}\right)^n \right],
\end{align}
where $T_r$ is a reference temperature, which is constrained to be larger than $\mathcal{O}(10)$ MeV from BBN \cite{DEramo:2017gpl}, and which corresponds to the temperature at which the energy density of $\phi$ equals that of radiation. In this study, we choose $T_r = 50$ MeV.

The $\phi$ component drives the universe's expansion to faster rates. Under this scenario, the second Friedmann equation reads
\begin{align}
    \label{eq:second_friedmann_eq_phi}
    \dot{\rho}_R + \lambda(T)\rho_R \frac{\dot{a}}{a} = -\dot{\rho_v},
\end{align}
where $\lambda(T)$ is a temperature-dependent coefficient, arising from the fact that $\phi$ has a different  equation of state than radiation \footnote{Specifically, $P_\phi=(n+1)\rho_\phi/3$.}. Assuming, as is natural to do, that Eq.(\ref{eq:rho_R_with_T}) is a good approximation when $\rho_v$ is turned on, then $\lambda(T)$ reads
\begin{align}
    \label{eq:lam_Trelation}
    \lambda(T)=4+n\frac{T^n}{T_r^n+T^n}.
\end{align}
Since the critical temperature $T_c$ is, for our choice of parameters, much larger than $T_r$, $\lambda(T)$ will decrease from $4+n$ to $4$ when the temperature drops below $T_r$. We do not anticipate any effect on our results above if $T_c\ll T_r$, as PBH only form at and below $T=T_c$.

\begin{figure}[t]
  \centering
\includegraphics[width=0.5\textwidth]{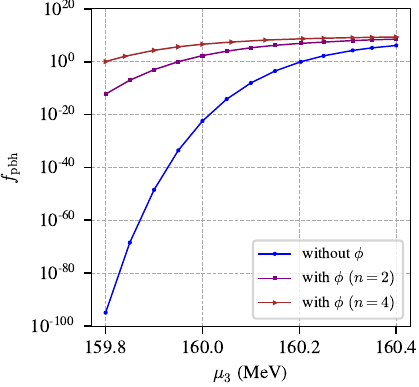}%
  \caption{$f_\text{pbh}$ with $\mu_3$ under $m_\phi=300$ MeV, $\omega=860$ MeV, and $c=0.14$ for different cosmological scenarios.
}
\label{fig:f_pbh_with_phi_components_research}
\end{figure}

Using the setup described above, we compute the $f_\text{pbh}$ with a $\phi$ component that red-shifts with $n=2$ and $n=4$, respectively.  We present our results in Fig.\ref{fig:f_pbh_with_phi_components_research}. 

We find that the PBH relic abundance is {\it enhanced} and its dependence on model parameters is weakened, and thus less fine-tuned. Interestingly, we also find that the super-exponential dependence still holds. This is not unexpected given  the analytic expression for $f_\text{pbh}$ and the definition of $A(T)$ and $\mathcal{Q}$ in Eq.(\ref{eq:A_T_definition},\ref{eq:Q_definition_new}): We first note that $T_p$ and $S_3(T_p)/T_p$ only depend on thermodynamics (what we referred to as model parameters), which stay invariant for different cosmological components; $a_\text{del}(T_p)$ also is invariant, given the assumption of thermal equilibrium  at all temperatures; the numerical results indicate that the large constant $\mathcal{N}(t_\text{pbh})$ is not sensitive to the addition of $\phi$ component. As a result, the only quantity in Eq.(\ref{eq:Q_definition_new}) that receives large modification is $H_\text{del}(T_p)$, which increases after introducing the $\phi$ component; larger values of $n$ naturally lead to larger values of $H_\text{del}(T_p)$. We conclude that a super-fast universe expansion will decrease the value of $\mathcal{Q}$ of order $\mathcal{O}(10)$, leading to a considerable, exponential  enhancement of $f_\text{pbh}$, as shown in Fig.\ref{fig:f_pbh_with_phi_components_research}. 

\section{Conclusions} \label{sec:conclusions}
We demonstrated that the production of PBHs from delayed vacuum decay obeys a {\it super-exponential} dependence on $S_3/T$ at its minimal temperature:
\begin{align}
    f_{\text{pbh}}\simeq \mathcal{M} \exp(-\mathcal{Q} \exp(-S_3(T_p)/T_p)),
\end{align}
where $f_{\text{pbh}}$ is the ratio of PBH-to-dark matter relative abundance; $\mathcal{M}$ and $\mathcal{Q}$ are model-dependent parameters; and $T_p$ is the peak temperature when $S_3(T)/T$ reaches its minimal value. $f_\text{pbh}$ is overwhelmingly more sensitive to variations in $S_3(T_p)/T_p$ than to comparable variations in $\mathcal{M}$ or $\mathcal{Q}$. It is thus the quantity $S_3(T_p)/T_p$ that {\it super-exponentially} controls the  abundance of the produced PBH. 

In the simplified model we consider, $\mathcal{M}\sim 4\times 10^7$ and $\mathcal{Q}\sim 1\times 10^{77}$. A very small variation in $S_3(T_p)/T_p$, arising from small changes in the model parameters, thus results in huge effects in $f_{\text{pbh}}$. 

Finally, we investigated the influence on  $f_\text{pbh}$ of a possible additional energy density  component $\phi$ that would dominate the universes' energy budget at early times, driving a non-standard, super-fast expansion rate. The $\phi$ component  reduces the value of the $\mathcal{Q}$ parameter in $f_\text{pbh}$, leading to an enhancement of the PBH relic abundance, and to a weaker dependence of $f_\text{pbh}$ on model parameters.

Future work will target more complex, richer models, possibly including multiple field directions, or multi-stage phase transitions, with and without a standard cosmology at early times.

\section*{Acknowledgements} \label{sec:acknowledgements}
We thank Michael Ramsey-Musolf and Tuomas V. I. Tenkanen for helpful discussions and the anonymous Referee for valuable suggestions and comments.
This work is partly supported by the U.S.\ Department of Energy grant number de-sc0010107 (SP).

\bibliography{biblio} 

\begin{thebibliography}{18}%
\makeatletter
\providecommand \@ifxundefined [1]{%
 \@ifx{#1\undefined}
}%
\providecommand \@ifnum [1]{%
 \ifnum #1\expandafter \@firstoftwo
 \else \expandafter \@secondoftwo
 \fi
}%
\providecommand \@ifx [1]{%
 \ifx #1\expandafter \@firstoftwo
 \else \expandafter \@secondoftwo
 \fi
}%
\providecommand \natexlab [1]{#1}%
\providecommand \enquote  [1]{``#1''}%
\providecommand \bibnamefont  [1]{#1}%
\providecommand \bibfnamefont [1]{#1}%
\providecommand \citenamefont [1]{#1}%
\providecommand \href@noop [0]{\@secondoftwo}%
\providecommand \href [0]{\begingroup \@sanitize@url \@href}%
\providecommand \@href[1]{\@@startlink{#1}\@@href}%
\providecommand \@@href[1]{\endgroup#1\@@endlink}%
\providecommand \@sanitize@url [0]{\catcode `\\12\catcode `\$12\catcode `\&12\catcode `\#12\catcode `\^12\catcode `\_12\catcode `\%12\relax}%
\providecommand \@@startlink[1]{}%
\providecommand \@@endlink[0]{}%
\providecommand \url  [0]{\begingroup\@sanitize@url \@url }%
\providecommand \@url [1]{\endgroup\@href {#1}{\urlprefix }}%
\providecommand \urlprefix  [0]{URL }%
\providecommand \Eprint [0]{\href }%
\providecommand \doibase [0]{http://dx.doi.org/}%
\providecommand \selectlanguage [0]{\@gobble}%
\providecommand \bibinfo  [0]{\@secondoftwo}%
\providecommand \bibfield  [0]{\@secondoftwo}%
\providecommand \translation [1]{[#1]}%
\providecommand \BibitemOpen [0]{}%
\providecommand \bibitemStop [0]{}%
\providecommand \bibitemNoStop [0]{.\EOS\space}%
\providecommand \EOS [0]{\spacefactor3000\relax}%
\providecommand \BibitemShut  [1]{\csname bibitem#1\endcsname}%
\let\auto@bib@innerbib\@empty
\bibitem [{\citenamefont {Escriv\`a}\ \emph {et~al.}(2022)\citenamefont {Escriv\`a}, \citenamefont {Kuhnel},\ and\ \citenamefont {Tada}}]{escriva}%
  \BibitemOpen
  \bibfield  {author} {\bibinfo {author} {\bibfnamefont {A.}~\bibnamefont {Escriv\`a}}, \bibinfo {author} {\bibfnamefont {F.}~\bibnamefont {Kuhnel}}, \ and\ \bibinfo {author} {\bibfnamefont {Y.}~\bibnamefont {Tada}},\ }\href {\doibase 10.1016/B978-0-32-395636-9.00012-8} {\  (\bibinfo {year} {2022}),\ 10.1016/B978-0-32-395636-9.00012-8},\ \Eprint {http://arxiv.org/abs/2211.05767} {arXiv:2211.05767 [astro-ph.CO]} \BibitemShut {NoStop}%
\bibitem [{\citenamefont {Carr}\ \emph {et~al.}(2021)\citenamefont {Carr}, \citenamefont {Kohri}, \citenamefont {Sendouda},\ and\ \citenamefont {Yokoyama}}]{carr}%
  \BibitemOpen
  \bibfield  {author} {\bibinfo {author} {\bibfnamefont {B.}~\bibnamefont {Carr}}, \bibinfo {author} {\bibfnamefont {K.}~\bibnamefont {Kohri}}, \bibinfo {author} {\bibfnamefont {Y.}~\bibnamefont {Sendouda}}, \ and\ \bibinfo {author} {\bibfnamefont {J.}~\bibnamefont {Yokoyama}},\ }\href {\doibase 10.1088/1361-6633/ac1e31} {\bibfield  {journal} {\bibinfo  {journal} {Rept. Prog. Phys.}\ }\textbf {\bibinfo {volume} {84}},\ \bibinfo {pages} {116902} (\bibinfo {year} {2021})},\ \Eprint {http://arxiv.org/abs/2002.12778} {arXiv:2002.12778 [astro-ph.CO]} \BibitemShut {NoStop}%
\bibitem [{\citenamefont {Liu}\ \emph {et~al.}(2022)\citenamefont {Liu}, \citenamefont {Bian}, \citenamefont {Cai}, \citenamefont {Guo},\ and\ \citenamefont {Wang}}]{Liu:2021svg}%
  \BibitemOpen
  \bibfield  {author} {\bibinfo {author} {\bibfnamefont {J.}~\bibnamefont {Liu}}, \bibinfo {author} {\bibfnamefont {L.}~\bibnamefont {Bian}}, \bibinfo {author} {\bibfnamefont {R.-G.}\ \bibnamefont {Cai}}, \bibinfo {author} {\bibfnamefont {Z.-K.}\ \bibnamefont {Guo}}, \ and\ \bibinfo {author} {\bibfnamefont {S.-J.}\ \bibnamefont {Wang}},\ }\href {\doibase 10.1103/PhysRevD.105.L021303} {\bibfield  {journal} {\bibinfo  {journal} {Phys. Rev. D}\ }\textbf {\bibinfo {volume} {105}},\ \bibinfo {pages} {L021303} (\bibinfo {year} {2022})},\ \Eprint {http://arxiv.org/abs/2106.05637} {arXiv:2106.05637 [astro-ph.CO]} \BibitemShut {NoStop}%
\bibitem [{\citenamefont {Kawana}\ \emph {et~al.}(2022)\citenamefont {Kawana}, \citenamefont {Kim},\ and\ \citenamefont {Lu}}]{ref1}%
  \BibitemOpen
  \bibfield  {author} {\bibinfo {author} {\bibfnamefont {K.}~\bibnamefont {Kawana}}, \bibinfo {author} {\bibfnamefont {T.}~\bibnamefont {Kim}}, \ and\ \bibinfo {author} {\bibfnamefont {P.}~\bibnamefont {Lu}},\ }\href {\doibase 10.1103/PhysRevD.108.103531} {\enquote {\bibinfo {title} {Pbh formation from overdensities in delayed vacuum transitions},}\ } (\bibinfo {year} {2022})\BibitemShut {NoStop}%
\bibitem [{\citenamefont {Flores}\ \emph {et~al.}(2024)\citenamefont {Flores}, \citenamefont {Kusenko},\ and\ \citenamefont {Sasaki}}]{ref2}%
  \BibitemOpen
  \bibfield  {author} {\bibinfo {author} {\bibfnamefont {M.~M.}\ \bibnamefont {Flores}}, \bibinfo {author} {\bibfnamefont {A.}~\bibnamefont {Kusenko}}, \ and\ \bibinfo {author} {\bibfnamefont {M.}~\bibnamefont {Sasaki}},\ }\href {\doibase 10.1103/physrevd.110.015005} {\enquote {\bibinfo {title} {Revisiting formation of primordial black holes in a supercooled first-order phase transition},}\ } (\bibinfo {year} {2024})\BibitemShut {NoStop}%
\bibitem [{\citenamefont {Jedamzik}\ and\ \citenamefont {Niemeyer}(1999)}]{ref5}%
  \BibitemOpen
  \bibfield  {author} {\bibinfo {author} {\bibfnamefont {K.}~\bibnamefont {Jedamzik}}\ and\ \bibinfo {author} {\bibfnamefont {J.}~\bibnamefont {Niemeyer}},\ }\href {\doibase 10.1103/PhysRevD.59.124014} {\enquote {\bibinfo {title} {Primordial black hole formation during first-order phase transitions},}\ } (\bibinfo {year} {1999})\BibitemShut {NoStop}%
\bibitem [{\citenamefont {Khlopov}\ \emph {et~al.}(1998)\citenamefont {Khlopov}, \citenamefont {Konoplich}, \citenamefont {Rubin},\ and\ \citenamefont {Sakharov}}]{Khlopov:1998nm}%
  \BibitemOpen
  \bibfield  {author} {\bibinfo {author} {\bibfnamefont {M.~Y.}\ \bibnamefont {Khlopov}}, \bibinfo {author} {\bibfnamefont {R.~V.}\ \bibnamefont {Konoplich}}, \bibinfo {author} {\bibfnamefont {S.~G.}\ \bibnamefont {Rubin}}, \ and\ \bibinfo {author} {\bibfnamefont {A.~S.}\ \bibnamefont {Sakharov}},\ }\href@noop {} {\  (\bibinfo {year} {1998})},\ \Eprint {http://arxiv.org/abs/hep-ph/9807343} {arXiv:hep-ph/9807343} \BibitemShut {NoStop}%
\bibitem [{\citenamefont {Gon\c{c}alves}\ \emph {et~al.}(2025)\citenamefont {Gon\c{c}alves}, \citenamefont {Kaladharan},\ and\ \citenamefont {Wu}}]{Goncalves:2024vkj}%
  \BibitemOpen
  \bibfield  {author} {\bibinfo {author} {\bibfnamefont {D.}~\bibnamefont {Gon\c{c}alves}}, \bibinfo {author} {\bibfnamefont {A.}~\bibnamefont {Kaladharan}}, \ and\ \bibinfo {author} {\bibfnamefont {Y.}~\bibnamefont {Wu}},\ }\href {\doibase 10.1103/PhysRevD.111.035009} {\bibfield  {journal} {\bibinfo  {journal} {Phys. Rev. D}\ }\textbf {\bibinfo {volume} {111}},\ \bibinfo {pages} {035009} (\bibinfo {year} {2025})},\ \Eprint {http://arxiv.org/abs/2406.07622} {arXiv:2406.07622 [hep-ph]} \BibitemShut {NoStop}%
\bibitem [{\citenamefont {Dai}\ \emph {et~al.}(2019)\citenamefont {Dai}, \citenamefont {Gregory},\ and\ \citenamefont {Stojkovic}}]{ref4}%
  \BibitemOpen
  \bibfield  {author} {\bibinfo {author} {\bibfnamefont {D.}~\bibnamefont {Dai}}, \bibinfo {author} {\bibfnamefont {R.}~\bibnamefont {Gregory}}, \ and\ \bibinfo {author} {\bibfnamefont {D.}~\bibnamefont {Stojkovic}},\ }\href {\doibase 10.1103/PHYSREVD.101.125012} {\enquote {\bibinfo {title} {Connecting the higgs potential and primordial black holes},}\ } (\bibinfo {year} {2019})\BibitemShut {NoStop}%
\bibitem [{\citenamefont {Wang}\ \emph {et~al.}(2020)\citenamefont {Wang}, \citenamefont {Huang},\ and\ \citenamefont {Zhang}}]{Wang:2020jrd}%
  \BibitemOpen
  \bibfield  {author} {\bibinfo {author} {\bibfnamefont {X.}~\bibnamefont {Wang}}, \bibinfo {author} {\bibfnamefont {F.~P.}\ \bibnamefont {Huang}}, \ and\ \bibinfo {author} {\bibfnamefont {X.}~\bibnamefont {Zhang}},\ }\href {\doibase 10.1088/1475-7516/2020/05/045} {\bibfield  {journal} {\bibinfo  {journal} {JCAP}\ }\textbf {\bibinfo {volume} {05}},\ \bibinfo {pages} {045} (\bibinfo {year} {2020})},\ \Eprint {http://arxiv.org/abs/2003.08892} {arXiv:2003.08892 [hep-ph]} \BibitemShut {NoStop}%
\bibitem [{\citenamefont {Khlopov}\ \emph {et~al.}(1999)\citenamefont {Khlopov}, \citenamefont {Konoplich}, \citenamefont {Rubin},\ and\ \citenamefont {Sakharov}}]{Khlopov:1999ys}%
  \BibitemOpen
  \bibfield  {author} {\bibinfo {author} {\bibfnamefont {M.~Y.}\ \bibnamefont {Khlopov}}, \bibinfo {author} {\bibfnamefont {R.~V.}\ \bibnamefont {Konoplich}}, \bibinfo {author} {\bibfnamefont {S.~G.}\ \bibnamefont {Rubin}}, \ and\ \bibinfo {author} {\bibfnamefont {A.~S.}\ \bibnamefont {Sakharov}},\ }\href@noop {} {\bibfield  {journal} {\bibinfo  {journal} {Grav. Cosmol.}\ }\textbf {\bibinfo {volume} {2}},\ \bibinfo {pages} {S1} (\bibinfo {year} {1999})},\ \Eprint {http://arxiv.org/abs/hep-ph/9912422} {arXiv:hep-ph/9912422} \BibitemShut {NoStop}%
\bibitem [{\citenamefont {Baker}\ \emph {et~al.}(2025)\citenamefont {Baker}, \citenamefont {Breitbach}, \citenamefont {Kopp},\ and\ \citenamefont {Mittnacht}}]{Baker:2021sno}%
  \BibitemOpen
  \bibfield  {author} {\bibinfo {author} {\bibfnamefont {M.~J.}\ \bibnamefont {Baker}}, \bibinfo {author} {\bibfnamefont {M.}~\bibnamefont {Breitbach}}, \bibinfo {author} {\bibfnamefont {J.}~\bibnamefont {Kopp}}, \ and\ \bibinfo {author} {\bibfnamefont {L.}~\bibnamefont {Mittnacht}},\ }\href {\doibase 10.1103/PhysRevD.111.063544} {\bibfield  {journal} {\bibinfo  {journal} {Phys. Rev. D}\ }\textbf {\bibinfo {volume} {111}},\ \bibinfo {pages} {063544} (\bibinfo {year} {2025})},\ \Eprint {http://arxiv.org/abs/2110.00005} {arXiv:2110.00005 [astro-ph.CO]} \BibitemShut {NoStop}%
\bibitem [{\citenamefont {Kanemura}\ \emph {et~al.}(2024)\citenamefont {Kanemura}, \citenamefont {Tanaka},\ and\ \citenamefont {Xie}}]{Kanemura:2024pae}%
  \BibitemOpen
  \bibfield  {author} {\bibinfo {author} {\bibfnamefont {S.}~\bibnamefont {Kanemura}}, \bibinfo {author} {\bibfnamefont {M.}~\bibnamefont {Tanaka}}, \ and\ \bibinfo {author} {\bibfnamefont {K.-P.}\ \bibnamefont {Xie}},\ }\href {\doibase 10.1007/JHEP06(2024)036} {\bibfield  {journal} {\bibinfo  {journal} {JHEP}\ }\textbf {\bibinfo {volume} {06}},\ \bibinfo {pages} {036} (\bibinfo {year} {2024})},\ \Eprint {http://arxiv.org/abs/2404.00646} {arXiv:2404.00646 [hep-ph]} \BibitemShut {NoStop}%
\bibitem [{\citenamefont {Callan}\ and\ \citenamefont {Coleman}(1977)}]{Callan:1977pt}%
  \BibitemOpen
  \bibfield  {author} {\bibinfo {author} {\bibfnamefont {C.~G.}\ \bibnamefont {Callan}, \bibfnamefont {Jr.}}\ and\ \bibinfo {author} {\bibfnamefont {S.~R.}\ \bibnamefont {Coleman}},\ }\href {\doibase 10.1103/PhysRevD.16.1762} {\bibfield  {journal} {\bibinfo  {journal} {Phys. Rev. D}\ }\textbf {\bibinfo {volume} {16}},\ \bibinfo {pages} {1762} (\bibinfo {year} {1977})}\BibitemShut {NoStop}%
\bibitem [{\citenamefont {Coleman}\ and\ \citenamefont {De~Luccia}(1980)}]{Coleman:1980aw}%
  \BibitemOpen
  \bibfield  {author} {\bibinfo {author} {\bibfnamefont {S.~R.}\ \bibnamefont {Coleman}}\ and\ \bibinfo {author} {\bibfnamefont {F.}~\bibnamefont {De~Luccia}},\ }\href {\doibase 10.1103/PhysRevD.21.3305} {\bibfield  {journal} {\bibinfo  {journal} {Phys. Rev. D}\ }\textbf {\bibinfo {volume} {21}},\ \bibinfo {pages} {3305} (\bibinfo {year} {1980})}\BibitemShut {NoStop}%
\bibitem [{\citenamefont {Cai}\ \emph {et~al.}(2024)\citenamefont {Cai}, \citenamefont {Hao},\ and\ \citenamefont {Wang}}]{Cai:2024nln}%
  \BibitemOpen
  \bibfield  {author} {\bibinfo {author} {\bibfnamefont {R.-G.}\ \bibnamefont {Cai}}, \bibinfo {author} {\bibfnamefont {Y.-S.}\ \bibnamefont {Hao}}, \ and\ \bibinfo {author} {\bibfnamefont {S.-J.}\ \bibnamefont {Wang}},\ }\href {\doibase 10.1007/s11433-024-2416-3} {\bibfield  {journal} {\bibinfo  {journal} {Sci. China Phys. Mech. Astron.}\ }\textbf {\bibinfo {volume} {67}},\ \bibinfo {pages} {290411} (\bibinfo {year} {2024})},\ \Eprint {http://arxiv.org/abs/2404.06506} {arXiv:2404.06506 [astro-ph.CO]} \BibitemShut {NoStop}%
\bibitem [{\citenamefont {Navas}\ \emph {et~al.}(2024)\citenamefont {Navas} \emph {et~al.}}]{ParticleDataGroup:2024cfk}%
  \BibitemOpen
  \bibfield  {author} {\bibinfo {author} {\bibfnamefont {S.}~\bibnamefont {Navas}} \emph {et~al.} (\bibinfo {collaboration} {Particle Data Group}),\ }\href {\doibase 10.1103/PhysRevD.110.030001} {\bibfield  {journal} {\bibinfo  {journal} {Phys. Rev. D}\ }\textbf {\bibinfo {volume} {110}},\ \bibinfo {pages} {030001} (\bibinfo {year} {2024})}\BibitemShut {NoStop}%
\bibitem [{\citenamefont {D'Eramo}\ \emph {et~al.}(2017)\citenamefont {D'Eramo}, \citenamefont {Fernandez},\ and\ \citenamefont {Profumo}}]{DEramo:2017gpl}%
  \BibitemOpen
  \bibfield  {author} {\bibinfo {author} {\bibfnamefont {F.}~\bibnamefont {D'Eramo}}, \bibinfo {author} {\bibfnamefont {N.}~\bibnamefont {Fernandez}}, \ and\ \bibinfo {author} {\bibfnamefont {S.}~\bibnamefont {Profumo}},\ }\href {\doibase 10.1088/1475-7516/2017/05/012} {\bibfield  {journal} {\bibinfo  {journal} {JCAP}\ }\textbf {\bibinfo {volume} {05}},\ \bibinfo {pages} {012} (\bibinfo {year} {2017})},\ \Eprint {http://arxiv.org/abs/1703.04793} {arXiv:1703.04793 [hep-ph]} \BibitemShut {NoStop}%
\end{thebibliography}%

\appendix*   
\section{Details of the simplified model} 
The toy model we consider consists of a single real scalar field, with an effective potential at finite temperature 
\begin{equation}
 \begin{aligned}
V(\phi,T)&=\frac{1}{2}\left( \frac{\mu_3 \omega - m_\phi^2}{2}+c T^2\right)\phi^2 -\frac{\mu_3}{3}\phi^3 \\
&\quad + \frac{m_\phi^2}{8\omega^2} \left( 1+\frac{\mu_3 \omega}{m_\phi^2}\right)\phi^4,
\end{aligned}   
\end{equation}
where $\mu_3$ is constrained in the range
\begin{align} m_\phi^2/\omega<\mu_3<3m_\phi^2/\omega.
\end{align}
The nucleation rate is given by
\begin{align}
    \Gamma(T)\simeq T^4 \left(\frac{S_3(T)}{2\pi T}\right)^{3/2}e^{-S_3(T)/T},
\end{align}
where the expression of $S_3(T)/T$ is given by the semi-analytical expression
\begin{align}
\frac{S_3(T)}{T}\simeq \frac{123.48 (\mu^2+c T^2)^{3/2}}{2^{3/2}T\mu_3^2}f\left(\frac{9(\mu^2+c T^2)\lambda}{2\mu_3^2} \right),
\end{align}
with
\begin{align}
\mu^2=\frac{\mu_3 \omega - m_\phi^2}{2},\quad \lambda=\frac{\mu_3 \omega + m_\phi^2}{2 \omega^2},
\end{align}
and
\begin{align}
    f(\mu)=1+\frac{\mu}{4}\left(1+\frac{2.4}{1-\mu}+\frac{0.26}{(1-\mu)^2}\right).
\end{align}
We note that the parameterization above is very accurate in the interval $\mu \in [0,1]$.

\end{document}